\documentclass[prl,showpacs,twocolumn,superscriptaddress,floatfix,amsmath]{revtex4}
\usepackage[latin1]{inputenc}
\usepackage[english]{babel}
\usepackage{graphicx,psfrag}
\usepackage{amsmath}
\usepackage{amssymb}
\usepackage{amscd}
\usepackage{eucal}
\usepackage{color}
\usepackage{bm}

\input{epsf}
\begin{document}
\title{Mesoscopic Spin Hall Effect}
\author{J.~H.~Bardarson}
\affiliation{Instituut-Lorentz, Universiteit Leiden, P.O. Box 9506,
2300 RA Leiden, The Netherlands}
\author{\.I.~Adagideli}
\affiliation{Institut f\"ur Theoretische Physik, Universit\"at Regensburg, D-93040, Germany}
\author{Ph.~Jacquod}
\affiliation{Physics Department,
   University of Arizona, 1118 E. 4$^{\rm th}$ Street, Tucson, AZ 85721, USA}
\date{October 2006}
\begin{abstract}
We investigate the spin Hall effect in ballistic chaotic quantum dots with
spin-orbit coupling. We show that a longitudinal charge current
can generate a pure transverse spin current.
While this transverse spin current is generically nonzero for a fixed sample, we show
that when the spin-orbit coupling time is large
compared to the mean dwell time inside the dot,
it fluctuates universally from sample to sample or upon variation
of the chemical potential with a vanishing average.
For a fixed sample configuration, the transverse spin current has a finite typical
value $\simeq e^2 V/h$, proportional to the longitudinal bias $V$
on the sample, and corresponding to about one excess open channel for one of
the two spin species. Our analytical results are in agreement with numerical results in a
diffusive system [W. Ren {\it et al}., Phys. Rev. Lett. {\bf 97}, 066603 (2006)] and are further confirmed by numerical simulation
in a chaotic cavity.
\end{abstract}
\maketitle{}

{\bf Introduction.}
The novel and rapidly expanding field of spintronics is interested in the
creation, manipulation, and detection of polarized or pure spin
currents~\cite{spintronics}.
The conventional methods of
doing spintronics are to use magnetic fields and/or
ferromagnets as parts of the creation-manipulation-detection cycle, and to use
the Zeeman coupling and the
ferromagnetic-exchange interactions to induce the spin dependency
of transport.
More recently, ways to generate spin accumulations and
spin currents based on the coupling of spin and orbital
degrees of freedom have been explored. Among these proposals, much
attention has been focused on the spin Hall effect (SHE), where pure spin
currents are generated by applied electric currents on spin-orbit (SO)
coupled systems.
Originally proposed by Dyakonov and Perel~\cite{DP71}, the idea was
resurrected by Hirsch~\cite{Hirsch} and extended to crystal SO field (the
intrinsic SHE) by Sinova {\it et al}.~\cite{macdonald} and Murakami
{\it et al}.~\cite{murakami}. The current agreement is
that the SHE vanishes for bulk, $k$-linear SO coupling for diffusive
two-dimensional electrons~\cite{Inoue04,halperin,macdonald2}. This result
is however specific to these systems~\cite{inanc}, and the SHE does not vanish for
impurity-generated SO coupling, two-dimensional hole systems
with either Rashba or Dresselhaus SO coupling, and for
finite-sized electronic systems~\cite{halperin,inanc}.
These predictions have been, to some
extent, confirmed by experimental observations of edge spin accumulations
in electron~\cite{Kato} and hole~\cite{Wunderlich} systems, and electrical
detection of spin currents via ferromagnetic
leads~\cite{Valenzuela}.

Most investigations of the SHE to date focused on disordered
conductors with spin-orbit interaction, where the disorder-averaged
spin Hall conductivity was calculated using either the Kubo
formalism or a diffusion equation approach~\cite{Hirsch,macdonald,macdonald2,loss,Inoue04,halperin,raimondi,murakami,inanc}.
Few numerical works alternatively used the scattering approach to
transport~\cite{markus} to calculate the average spin Hall
conductance of explicitly finite-sized samples connected to external
electrodes. These investigations were however restricted to
tight-binding Hamiltonians with no or weak disorder in simple
geometries~\cite{branislav,sinova,sheng}. The data of
Ref.~\cite{guo} in particular suggest that diffusive samples with
large enough SO coupling exhibit universal fluctuations of the spin
Hall conductance $G_{\text{sH}}$ with ${\rm rms}[G_{\text{sH}}]
\approx 0.18 e/4\pi$. These numerical investigations call for an
analytical theory of the SHE in mesoscopic systems.
It is the purpose of this article to provide such a theory.

We analytically investigate the DC spin Hall effect in mesoscopic
cavities with SO coupling. We calculate both the ensemble-average
and the fluctuations of the transverse spin current generated by a
longitudinal charge current. Our approach is based on random matrix
theory (RMT)~\cite{carlormp}, and is valid for ballistic chaotic
and mesoscopic diffusive systems at low temperature, in the limit
when the spin-orbit coupling time is much shorter than the mean
dwell time of the electrons in the cavity, $\tau_{\rm so} \ll
\tau_{\rm dwell}$ \cite{caveat1}. We show that while the transverse
spin current is generically nonzero for a typical sample, its sign
and amplitude fluctuate universally, from sample to sample or upon
variation of the chemical potential with a vanishing average. We
find that for a typical ballistic chaotic quantum dot, the
transverse spin current corresponds to slightly less than one excess
open channel for one of the two spin species. These analytical
results are confirmed by numerical simulations for a stroboscopic
model of a ballistic chaotic cavity.

\begin{figure}
\includegraphics[width=0.73\columnwidth]{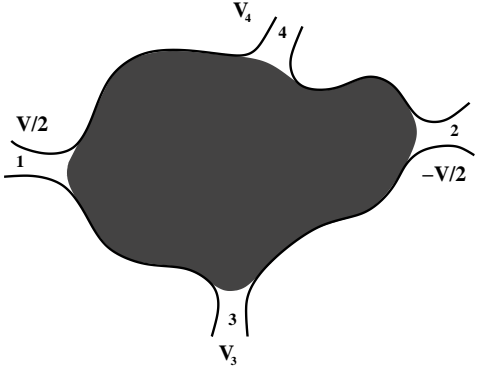}
\vspace{-0.3cm}
\caption{\label{fig1} Ballistic quantum dot connected to four electrodes.
The longitudinal
bias $V$ induces a charge current through terminals 1 and 2, while the voltages $V_{3,4}$ are adjusted such that no charge current flows through the
transverse leads 3 and 4. Spin-orbit coupling is active only in the gray
region. }
\end{figure}

{\bf Scattering approach.} We consider a ballistic chaotic quantum dot coupled to four external electrodes
via ideal point contacts, each with $N_i$ open channels ($i=1,\ldots 4$).
The geometry is sketched in Fig.~\ref{fig1}.
Spin-orbit coupling exists only inside the dot, and
the electrochemical potentials in the
electrodes are spin-independent. A bias voltage $V$ is
applied between the longitudinal electrodes labeled 1 and 2.
The voltages $V_3$ and $V_4$ are set such that no net charge current flows
through the transverse electrodes 3 and 4. We will focus on
the magnitude of the spin current through electrodes 3 and 4, in the limit
when the openings to the electrodes are small enough, and the spin-orbit
coupling strong enough that $\tau_{\rm so} \ll \tau_{\rm dwell}$.

We write the spin-resolved current through the $i$-th electrode
as \cite{markus}
\begin{equation}\label{scatt0}
I_i^\sigma = \frac{e^2}{h} \sum_{j,\sigma'} T^{\sigma,\sigma'}_{ij} (V_i-V_j).
\end{equation}
The spin-dependent transmission coefficients are obtained by summing
over electrode channels
\begin{eqnarray}
T^{\sigma,\sigma'}_{i,j} & = & \sum_{m \in i} \sum_{n \in j}
|t_{m,\sigma;n,\sigma'}|^2,
\end{eqnarray}
i.e. $t_{m,\sigma;n,\sigma'}$ is the transmission amplitude
for an electron initially in a spin state $\sigma'$ in channel $n$ of electrode $j$
to a spin state $\sigma$ in channel $m$ of electrode $i$. The transmission amplitudes $t$ are the elements of the $2N_T\times2N_T$
scattering matrix $S$, with $N_T = \sum_{i=1}^4N_i$.

We are interested in the transverse spin currents
$I_{i}^{(z)} = I_{i}^{\uparrow}-I_{i}^{\downarrow} $, $i=3,4$, under the two constraints that
(i) charge current vanishes in the transverse leads, $I_{i}^{\uparrow}+I_{i}^{\downarrow}=0$, $i=3,4$
and (ii) the charge current is conserved, $I_1=-I_2 = I$. From Eq.~(\ref{scatt0}),
transport through the system is then described by the following equation
\begin{widetext}
\begin{eqnarray}\label{scatt1}
\left(
\begin{array}{c}
2 J \\
J_3^{(z)} \\
J_4^{(z)}
\end{array}
\right) &=& \left(
\begin{array}{ccc}
2N_1 - {\mathcal T}_{11}^{(0)}+2N_2-{\mathcal T}_{22}^{(0)} + {\mathcal T}_{12}^{(0)}+
{\mathcal T}_{21}^{(0)} \; & {\mathcal T}_{23}^{(0)}
-{\mathcal T}_{13}^{(0)} \; & {\mathcal T}_{24}^{(0)}
-{\mathcal T}_{14}^{(0)} \\
 {\mathcal T}_{32}^{(z)}
-{\mathcal T}_{31}^{(z)} \; &  -{\mathcal T}_{33}^{(z)} \; &
-{\mathcal T}_{34}^{(z)} \\
 {\mathcal T}_{42}^{(z)}
-{\mathcal T}_{41}^{(z)} \; &  -{\mathcal T}_{43}^{(z)} \; &
-{\mathcal T}_{44}^{(z)}
\end{array}
\right)
\left(
\begin{array}{c}
1/2\\
\tilde{V}_3\\
\tilde{V}_4
\end{array}
\right),
\end{eqnarray}
\end{widetext}
where the transverse voltages (in units of $V$) read
\begin{subequations}
\begin{align}
\tilde{V}_3 &=   \frac{1}{2}\frac{{\mathcal T}_{34}^{(0)}({\mathcal T}_{42}^{(0)}-{\mathcal T}_{41}^{(0)})
+(2N_4-{\mathcal T}_{44}^{(0)})({\mathcal T}_{32}^{(0)}-{\mathcal T}_{31}^{(0)})}
{{\mathcal T}_{34}^{(0)} {\mathcal T}_{43}^{(0)}
-(2N_3-{\mathcal T}_{33}^{(0)}) (2N_4-
 {\mathcal T}_{34}^{(0)})  }, \\
\tilde{V}_4 &=  \frac{1}{2}\frac{{\mathcal T}_{43}^{(0)}({\mathcal T}_{32}^{(0)}-{\mathcal T}_{31}^{(0)})
+(2N_3-{\mathcal T}_{33}^{(0)})({\mathcal T}_{42}^{(0)}-{\mathcal T}_{41}^{(0)})}
{{\mathcal T}_{34}^{(0)} {\mathcal T}_{43}^{(0)}
-(2N_3-{\mathcal T}_{33}^{(0)}) (2N_4-
 {\mathcal T}_{34}^{(0)})  },
\end{align}
\end{subequations}
and we defined the dimensionless currents $I=e^2 V J/h$.
We introduced generalized transmission probabilities
\begin{equation}
\mathcal{T}_{ij}^{(\mu)} =\sum_{m \in i,n \in j}{\rm Tr} [ (t_{mn})^\dagger
\sigma^{(\mu)} t_{mn}], \;\;\;\; \mu = 0,x,y,z,
\end{equation}
where $\sigma^{(\mu)}$ are Pauli matrices ($\sigma^{(0)}$ is the identity matrix)
and one traces over the spin degree of freedom.

{\bf Random matrix theory.}
We calculate the average and fluctuations of the transverse spin currents
$J_i^{(\mu)}$, $\mu=x,y,z$ within the framework of RMT.
Accordingly, we replace
the scattering matrix $S$ by a random unitary matrix, which, in our
case of a system with time reversal symmetry (absence of magnetic field)
and totally broken spin rotational symmetry (strong spin-orbit coupling),
has to be taken from the circular symplectic ensemble
(CSE)~\cite{carlormp,caveat2,Ale01,Bro02}.
We rewrite the generalized transmission
probabilities $\mathcal{T}_{ij}^{(\mu)}$
as a trace over $S$
\begin{align}
  \label{eq:TfullTr}
  \mathcal{T}_{ij}^{(\mu)} &= \text{Tr}\,[Q_i^{(\mu)}SQ_j^{(0)}S^\dagger], \\
  [Q_i^{(\mu)}]_{m\alpha,n\beta} &= \notag
  \begin{cases}
    \delta_{mn}~\sigma^{(\mu)}_{\alpha\beta}, & \sum_{j=1}^{i-1}N_j < m \leq \sum_{j=1}^{i} N_j, \\
    0, & \text{otherwise}.
  \end{cases}
\end{align}
Here, $m$ and $n$ are channel indices, while
$\alpha$  and $\beta$ are spin indices. The trace is taken
over both set of indices.

Averages, variances, and covariances of the generalized transmission
probabilities~\eqref{eq:TfullTr} over the CSE can be calculated using the
method of Ref.~\cite{Bro96}. For the average transmission
probabilities, we find
\begin{equation}
  \label{eq:Tmean}
  \langle \mathcal{T}^{(\mu)}_{ij} \rangle = \frac{2\delta_{\mu 0}}{N_T-1/2}\left(N_iN_j - \frac{1}{2}N_i\delta_{ij}\right),
\end{equation}
while variances and covariances are given by
\begin{widetext}
\begin{align}
  \label{eq:Covb4}
  &\langle \delta \mathcal{T}_{ij}^{(\mu)} \delta \mathcal{T}_{kl}^{(\nu)} \rangle =
  \frac{4\delta_{\mu\nu}}{N_T(2N_T-1)^2(2N_T-3)}\Big\{ N_iN_j(N_T-1)(2N_T-1)(\delta_{ik}\delta_{jl} + \delta_{il}\delta_{jk}\delta_{\mu0})
   \notag \\
  &+ (N_iN_k\delta_{ij}\delta_{kl} - 2N_iN_kN_l\delta_{ij}
  - 2N_iN_jN_k\delta_{kl} + 4N_iN_jN_kN_l)\delta_{\mu0}
  - N_iN_T(2N_T-1)\delta_{ijkl} +(2N_T-1) \times\\
  &\Big[ N_iN_l\delta_{ijk} + N_iN_k\delta_{ijl}\delta_{\mu0}
  +N_iN_j(\delta_{ikl}+\delta_{jkl}\delta_{\mu0})
  - N_iN_jN_l(\delta_{ik}+\delta_{jk}\delta_{\mu0}) - N_iN_jN_k\delta_{\mu0}(\delta_{il}
  + \delta_{jl})\Big] \Big\}, \notag
\end{align}
\end{widetext}
where $\delta \mathcal{T}_{ij}^{(\mu)} = \mathcal{T}_{ij}^{(\mu)} - \langle\mathcal{T}_{ij}^{(\mu)}\rangle$.

Because the transverse potentials $\tilde{V}_{3,4}$ are spin-independent, they
are not correlated with $\mathcal{T}^{(\mu)}_{ij}$. Additionally taking
Eq.~(\ref{eq:Tmean}) into account, one concludes that the average transverse
spin current vanishes ($i=3,4$),
\begin{equation}
  \label{eq:Imean}
\langle J_i^{(\mu)} \rangle
= \frac{1}{2}
\langle {\mathcal T}_{i2}^{(\mu)} -{\mathcal T}_{i1}^{(\mu)} \rangle - \sum_{j=3,4}\langle {\mathcal T}_{ij}^{(\mu)}\rangle\langle \tilde{V}_j
\rangle = 0.
\end{equation}
However, for a given sample at a fixed chemical potential $J_i^{(\mu)}$
will in general be finite.
We thus calculate $\text{var}\;[J_i^{(\mu)}]$. We first note that
$\langle \tilde{V}_{3,4}\rangle/V =(N_1-N_2)/ 2(N_1+N_2)$, and
that $\text{var}\;[\tilde{V}_{3,4}]$
vanishes to leading order in the inverse number of channels. One thus has
\begin{align}
  &\text{var}\;[J_i^{(\mu)}] = \frac{1}{4}\sum_{j=1,2}\text{var}[\mathcal{T}_{ij}^{(\mu)}] -
  \frac{1}{2}\,\text{covar}[\mathcal{T}_{i1}^{(\mu)},\mathcal{T}_{i2}^{(\mu)}]   \\
  &+\sum_{j=3,4}\Big\{\text{var}[\mathcal{T}_{ij}^{(\mu)}]\langle \tilde{V}_j
  \rangle^2+\text{covar}[\mathcal{T}_{i1}^{(\mu)}-\mathcal{T}_{i2}^{(\mu)},\mathcal{T}_{ij}^{(\mu)}]\langle \tilde{V}_j \rangle \Big\} \notag\\
  &+2\, \text{covar}[\mathcal{T}_{i3}^{(\mu)},\mathcal{T}_{i4}^{(\mu)}]\langle \tilde{V}_3 \rangle\langle \tilde{V}_4 \rangle \notag.
\end{align}
From Eq.~\eqref{eq:Covb4} it follows that
\begin{equation}
  \label{eq:Is}
  \text{var}\;[J_i^{(\mu)}] =\frac{4N_iN_1N_2(N_T-1)}{N_T(2N_T-1)(2N_T-3)(N_1+N_2)}.
\end{equation}
Eqs.~(\ref{eq:Imean}) and (\ref{eq:Is}) are our main results. They show
that, while the average transverse spin current vanishes, it
exhibits universal sample-to-sample fluctuations.
The origin of this universality is the same as for
charge transport \cite{carlormp}, and relies on the fact expressed in
Eq.~(\ref{eq:Covb4}) that to leading order,
spin-dependent transmission correlators do not scale with the
number of channels.
The spin current carried by a single typical sample is given by
${\rm rms}[J_i^{(\mu)}] \times e^2 V/h$, and is thus of order $e^2 V/h$
in the limit of large number of channels. In other words, for a given sample,
one spin species has of order one more open transport channel than
the other one.
For a fully symmetric configuration, $N_i \equiv N$, the spin
current fluctuates universally for large $N$, with
${\rm rms}[I_3^{z}] \simeq (e^2 V/h)/\sqrt{32}$. This
translates into universal fluctuations of the transverse spin conductance with
${\rm rms} [G_{\rm sH}] = (e/4 \pi \sqrt{32})\approx 0.18 (e/4 \pi)$ in
agreement with Ref.~\cite{guo}.

{\bf Numerical simulation.}
In the diffusive setup of
Ref.~\cite{guo} the universal regime is not very large and thus it is
difficult to unambiguously identify it. We therefore present numerical
simulations in chaotic cavities to further illustrate our analytical
predictions~\eqref{eq:Imean} and~\eqref{eq:Is}.

We model the
electronic dynamics inside a chaotic ballistic cavity by the spin kicked
rotator~\cite{Sch89,Bar05}, a one-dimensional quantum map which
we represent by a $2M \times 2M$ Floquet (time-evolution) matrix~\cite{Izr90}
\begin{subequations}
\label{eq:floquet}
\begin{align}
\mathcal{F}_{ll'}&=(\Pi U X U^\dagger \Pi)_{ll'}, \quad l,l' = 0,1,\ldots, M-1,\\
\Pi_{ll'}&=\delta_{ll'}e^{-i\pi (l+l_0)^2/M }\sigma_0, \\
U_{ll'}&=M^{-1/2}e^{-i2\pi ll'/M}\sigma_0,\\
X_{ll'}&= \delta_{ll'}e^{-i(M/4\pi)V(2\pi l/M)}.
\end{align}
\end{subequations}
The matrix size $2 M=2 L/\lambda_F \gg 1$ is given by twice
the ratio of the linear system size to the Fermi wavelength.
The matrix $X$ with
\begin{equation}
V(p) =  K\cos p\,\sigma_0 + K_\text{so}(\sigma_x\sin2p
+ \sigma_z\sin p).
\end{equation}
corresponds to free SO coupled motion interrupted periodically by kicks described by the matrix $\Pi$, corresponding to scattering off the
boundaries of the quantum dot. In this form, the model is time-reversal symmetric, and the
parameter $l_0$ ensures that no additional symmetry exists in the system.
The map is classically chaotic for kicking strength $K \gtrsim 7.5$, and
$K_\text{so}$ is related to the SO coupling
time $\tau_\text{so}$ (in units of the stroboscopic period)
through
$\tau_\text{so} = 32\pi^2/K_\text{so}^2M^2$~\cite{Bar05}.
From (\ref{eq:floquet}), we construct the
quasienergy-dependent scattering matrix~\cite{fyodorov}
\begin{equation}
S(\varepsilon)=P[e^{-i\varepsilon}-{\mathcal F}(1-P^TP)]^{-1}{\mathcal F}P^T,
\end{equation}
with $P$ a $2N_T \times 2M$ projection matrix
\begin{equation}
 P_{k\alpha,k'\beta} =
\begin{cases}
    \delta_{\alpha\beta} &\text{if } k' = l^{(k)}, \\
    0 &\text{otherwise}.
\end{cases}
\end{equation}
The $l^{(k)}$ ($k=1,2,\ldots, 2N_T$, labels the modes) give
the position in phase space of the attached leads. The mean dwell
time $\tau_\text{dwell}$ (in units of the stroboscopic period) is given by
$\tau_\text{dwell} = M/N_T$. At large enough SO coupling, this model has been shown to exhibit
the universality of the CSE.
We refer the reader to Ref.~\cite{Bar05} for further details on the model.

\begin{figure}[tb]
  \begin{center}
    \includegraphics[width=0.75\columnwidth]{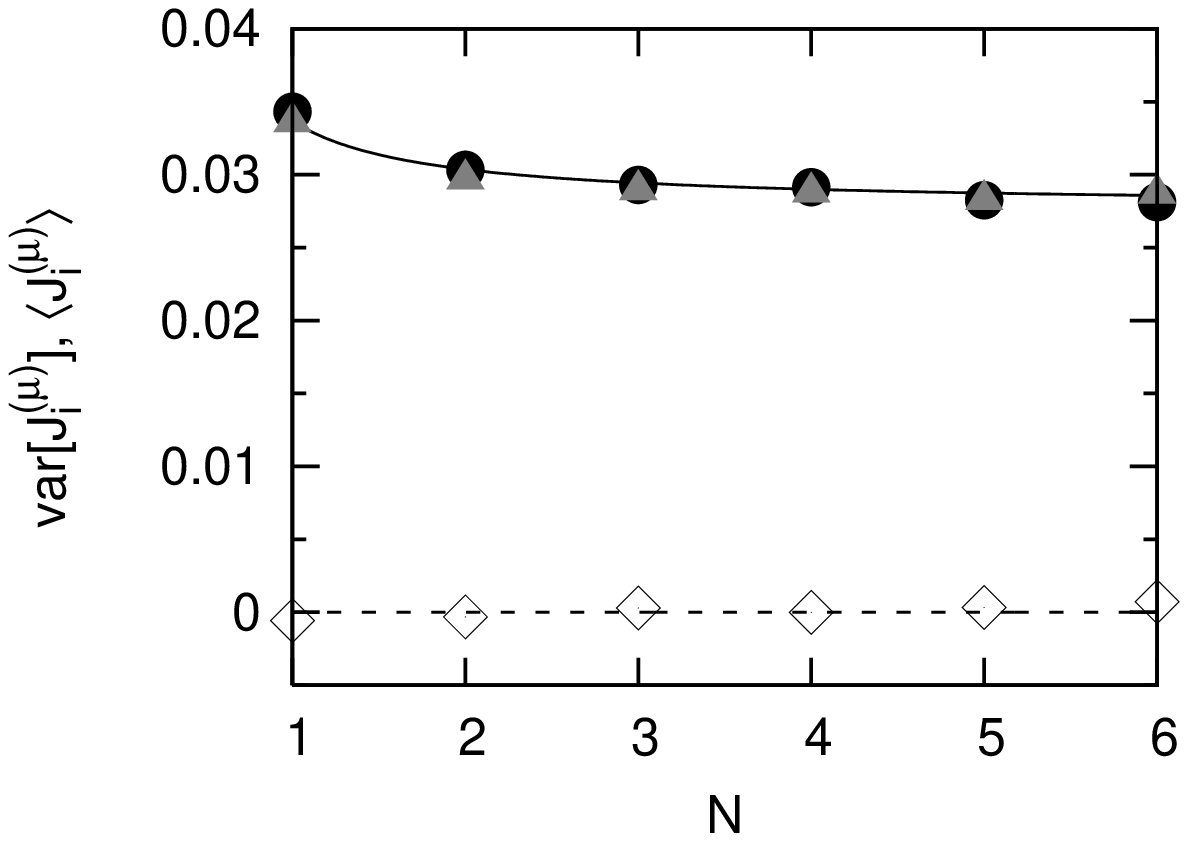} \\
    \vspace{-0.2cm}
    \includegraphics[width=0.75\columnwidth]{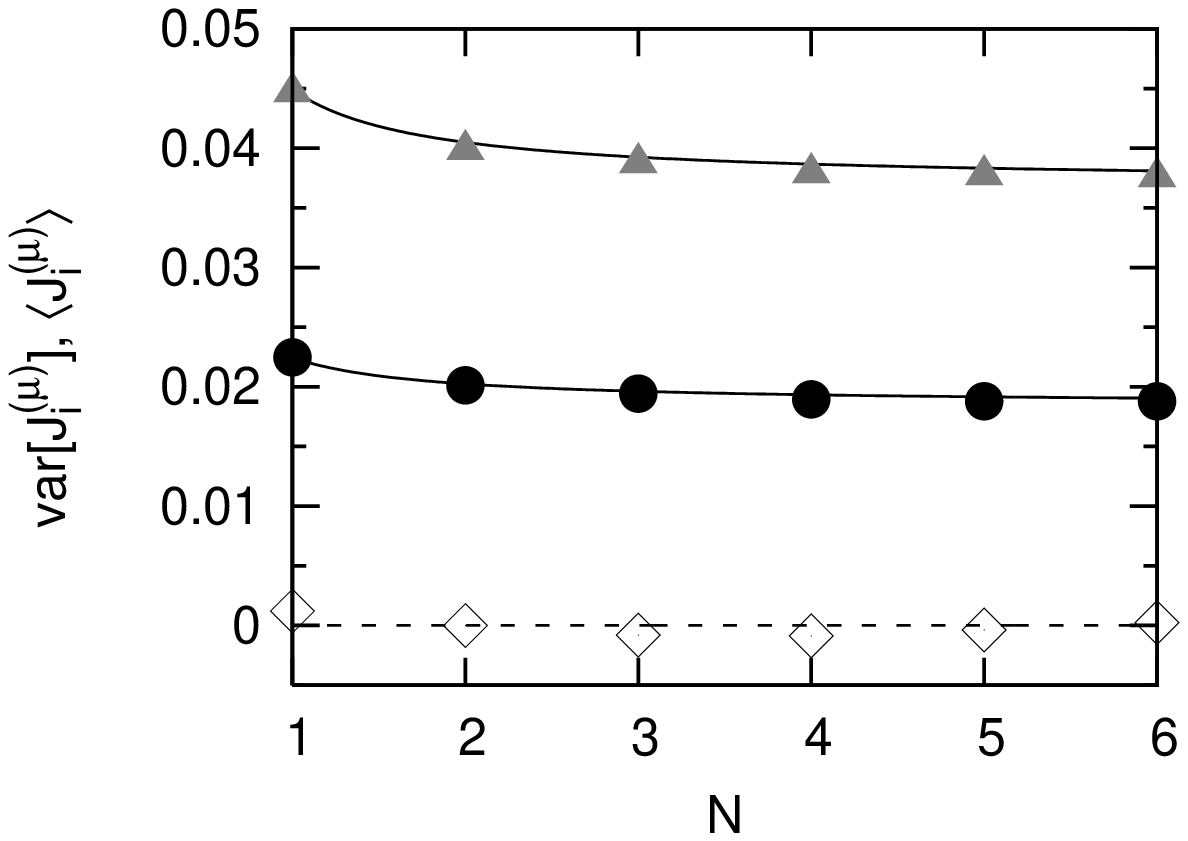}
  \end{center}
\vspace{-0.5cm}
  \caption{Average and variance of the transverse spin current vs. the
number of modes. Top panel: longitudinally symmetric configuration with
$N_1 = N_2 =  2N_3 = 2N_4 = 2N$; bottom panel: longitudinally asymmetric
configuration with $N_2
  = N_4 = 2N_1 = 2N_3 = 2N$. In both cases the total number of
modes $N_T = 6N$. The solid (dashed) lines give the analytical
  prediction~\eqref{eq:Imean} [\eqref{eq:Is}] for the mean (variance) of the spin currents. Empty diamonds correspond to
$\langle J_i^{(\mu)}\rangle$, circles to $\text{var}\;[J_3^{(\mu)}]$
and triangles to $\text{var}\;[J_4^{(\mu)}]$.}\label{fig:Is}
\end{figure}

Averages were performed over 35 values of $K$ in the range $41 < K < 48$,
25 values of $\varepsilon$ uniformly distributed in
$0 < \varepsilon < 2\pi$, and 10 different lead positions $l^{(k)}$.
We set the strength of $K_{\rm so}$
such that $\tau_\text{so} = \tau_\text{dwell}/1250$, and fixed values of
$M=640$ and $l_0 = 0.2$.

Our numerical results are presented in Fig.~\ref{fig:Is}.
Two cases were considered, the longitudinally
symmetric ($N_1 = N_2$) and asymmetric ($N_1 \neq N_2$) configurations.
In both cases, the numerical data fully confirm our predictions
that the average spin current vanishes
and that the variance of the transverse spin current is universal, i.e.
it does not depend on $N$ for large enough value of $N$.
In the asymmetric case $N_4 = 2N_3$, the
variance of the spin current in lead 4 is twice as big as in lead 3,
giving further confirmation to Eq.~(\ref{eq:Is}).

{\bf Conclusion.} We have calculated the average and mesoscopic fluctuations of the
transverse spin current generated by a charge current through a chaotic quantum dot with SO coupling.
We find that, from sample to sample, the spin current fluctuates
universally around zero average. In particular, for
a fully symmetric configuration $N_i \equiv N$, this translates into
universal fluctuations
of the spin conductance with ${\rm rms} [G_{\rm sH}] = (e/4 \pi
\sqrt{32})\approx 0.18 (e/4 \pi)$. This analytically
establishes the universality observed numerically in
Ref.~\cite{guo}.

We thank C.W.J.\ Beenakker for valuable comments on the manuscript.
JHB acknowledges support by the European Community's Marie Curie
Research Training Network under contract MRTN-CT-2003-504574,
Fundamentals of Nanoelectronics. IA acknowledges support by the
Deutsche Forschungsgemeinschaft within the cooperative research
center SFB 689 ``Spin phenomena in low dimensions'' and NSERC Canada
discovery grant number R8000.


\vspace{-0.6cm}
\begin{thebibliography}{99}
\vspace{-0.8cm}
\bibitem{spintronics} I. Zuti\'c {\it et al}., Rev. Mod. Phys. {\bf 76}, 323 (2004).
\bibitem{DP71} M.I. Dyakonov and V.I. Perel, Sov. Phys. JETP Lett. \textbf{13}, 467 (1971); Phys. Lett. A \textbf{35},459 (1971).
\bibitem{Hirsch} J. E. Hirsch, Phys. Rev. Lett. \textbf{83}, 1834 (1999)
\bibitem{macdonald} J. Sinova {\it et al}., Phys. Rev. Lett. {\bf 92}, 126603 (2004).
\bibitem{murakami} S. Murakami, Phys. Rev. B {\bf 69}, 241202(R) (2004).
\bibitem{Inoue04} J.-I. Inoue, G.E.W. Bauer, and L.W. Molenkamp,
Phys. Rev. B {\bf 70}, 041303(R) (2004).
\bibitem{halperin} E.G. Mishchenko, A.V. Shytov, and B.I. Halperin, Phys. Rev. Lett. {\bf 93},
226602 (2004).
\bibitem{macdonald2} A.A. Burkov, A.S. N\'u\~nez, and A.H. MacDonald, Phys. Rev. B
{\bf 70}, 155308 (2004).
\bibitem{inanc} I. Adagideli and G.E.W. Bauer, Phys. Rev. Lett. {\bf 95}, 256602 (2005).
\bibitem {Kato}
Y.K. Kato~\textit{et al}.,
Science \textbf{306}, 1910 (2004);
V. Sih~\textit{et al}.,
Nature Phys. 1, 31-35 (2005).
\bibitem{Wunderlich}
J. Wunderlich~\textit{et al}.,
Phys. Rev. Lett. \textbf{94}, 047204 (2005);
%
\bibitem {Valenzuela}
  E. Saitoh \textit{et al}., Appl. Phys. Lett. \textbf{88}, 182509 (2006);
  S.O.~Valenzuela and M. Tinkham, Nature \textbf{442}, 176 (2006);
  T.~Kimura \textit{et al.}, cond-mat/0609304.
\bibitem{loss} J. Schliemann and D. Loss, Phys. Rev. B {\bf 71}, 085308 (2005).
\bibitem{raimondi} R. Raimondi and P. Schwab, Phys. Rev. B {\bf 71}, 033311 (2005).
\bibitem{markus} M. B\"uttiker, Phys. Rev. Lett. {\bf 57}, 1761 (1986).
\bibitem{branislav} B.K. Nikoli\'c, L.P. Z\^arbo, and S. Souma, Phys. Rev. B
{\bf 72}, 75361 (2005).
\bibitem{sinova} E.M. Hankiewicz, L.W. Molenkamp, T. Jungwirth, and J. Sinova,
Phys. Rev. B {\bf 70}, 241301(R) (2004).
\bibitem{sheng} L. Sheng, D.N. Sheng, and C.S. Ting, Phys. Rev. Lett. {\bf 94}, 016602 (2005).
\bibitem{guo} W. Ren, Z. Qiao, J. Wang, Q. Sun, and H. Guo,
Phys. Rev. Lett. {\bf 97}, 066603 (2006).
\bibitem{carlormp}
C.~W.~J. Beenakker, Rev.\ Mod.\ Phys. \textbf{{69}}, {731} ({1997}).
\bibitem{caveat1} It is at present unclear to us whether the validity of RMT
for the spin Hall effect in mesoscopic diffusive samples requires
averaging over lead positions, in addition to disorder averaging.
\bibitem{caveat2} We assume that
the SO coupling parameters are sufficiently
nonuniform, so that SO cannot be removed from
the Hamiltonian by a gauge transformation, see~:
\cite{Ale01,Bro02}.
\bibitem{Ale01} I.~L. Aleiner and V.~I. {Fal'ko}, {Phys.\ Rev.\ Lett.} \textbf{{87}}, {256801} ({2001}).

\bibitem{Bro02} {P.~W.} Brouwer, {J.~N.~H.~J.} {Cremers}, {and} {B.~I.} {Halperin}, {Phys.\ Rev.\ B}
  \textbf{{65}}, {081302(R)} ({2002}).
\bibitem{Bro96}
{P.~W.} {Brouwer} {and} {C.~W.~J.} {Beenakker}, {J.\ Math.\ Phys.} \textbf{{37}}, {4904} ({1996}).

\bibitem{Sch89}
{R.}~{Scharf}, {J.\ Phys.\ A} \textbf{{22}}, {4223} ({1989}).

\bibitem{Bar05}
{J.~H.} {Bardarson}, {J.}~{Tworzyd{\l}o}, {and} {C.~W.~J.} {Beenakker}, {Phys.\ Rev.\ B}
  \textbf{{72}}, {235305} ({2005}).

\bibitem{Izr90}
{F.~M.~Izrailev}, {Phys.\ Rep.} \textbf{196}, 299 (1990).


\bibitem{fyodorov} Y.V. Fyodorov and H.-J. Sommers, JETP Lett. {\bf 72},
422 (2000).


\end{thebibliography}
\end{document}